\documentclass[aps,prl,reprint,letterpaper,showpacs]{revtex4-1}
\usepackage{keyval}%
\usepackage{graphicx}
\usepackage{dcolumn}
\usepackage{bm}
\usepackage{color}

\setkeys{Gin}{width=0.95\columnwidth,clip=true}

\newcommand{\ignore}[1]{}

\newcommand{\etal}{\textsl{et~al.}}

\begin{document}

\title{Unified Picture for Magnetic Correlations in Iron-Based Superconductors}
\author{Wei-Guo Yin}
\email[Corresponding author.\\ ]{wyin@bnl.gov}%
\author{Chi-Cheng Lee}
\author{Wei Ku}
\affiliation{Condensed Matter Physics and Materials Science
Department, Brookhaven National Laboratory, Upton, New York 11973,
USA
}

\date{\today}

\begin{abstract}
The varying metallic antiferromagnetic correlations observed in
iron-based superconductors are unified in a model consisting of both
itinerant electrons and localized spins. The decisive factor is
found to be the sensitive competition between the superexchange
antiferromagnetism and the orbital-degenerate double-exchange
ferromagnetism. Our results reveal the crucial role of Hund's rule
coupling for the strongly correlated nature of the system and
suggest that the iron-based superconductors are closer kin to
manganites than cuprates in terms of their diverse magnetism and
incoherent normal-state electron transport. This unified picture
would be instrumental for exploring other exotic properties and the
mechanism of superconductivity in this new class of superconductors.
\end{abstract}


\maketitle

Recently high-temperature superconductivity has been observed in a
number of doped iron-based layer materials near a static
antiferromagnetic (AF) order
\cite{syn:1111:Kamihara,syn:122:Rotter,syn:11:Yeh,neu:1111:Cruz,neu:122:Huang,neu:11:Li}
and with a spin resonance \cite{neu:122:Christianson,neu:11:Qiu}, a
pattern exhibited previously by the copper oxides. Intriguingly, in
contrast to the universal insulating checkerboard AF order in
undoped copper oxides, the AF orders in this new class of
superconductors are metallic and material-dependent: ``collinear''
(Fig.~1a) in undoped pnictides LaOFeAs and BaFe$_2$As$_2$
\cite{neu:1111:Cruz,neu:122:Huang} and ``bicollinear'' (Fig.~1b) in
undoped chalcogenide FeTe \cite{neu:11:Li}. This newly unveiled
magnetic diversity has greatly promoted the magnetic mechanism as a
general route to high-temperature superconductivity
\cite{review:Mazin}. It is thus essential to understand how these AF
correlations developed in the first place \cite{review:Lumsden}.

The fact that all the iron-based superconductors have similar
crystal structure, electronic structure, and Fermi-surface topology
\cite{arpes:1111:Lu,*arpes:122:Ding,*arpes:11:Xia} suggests that
their metallic magnetism has \emph{a common origin}. This is further
supported by the spin resonance in the superconducting state that
appears to be universally collinear like
\cite{neu:122:Christianson,neu:11:Qiu}. Moreover, it was shown
\cite{dft:11:Moon} that FeTe could switch from bicollinear to
collinear by decreasing the anion height from the Fe plane. These
observations call for a unified picture that hosts a sensitive
competition between the collinear and bicollinear AF orders.

However, previous model analyses did not reveal a close relationship
between these two orders. The collinear order has been widely noted
as a spin-density-wave state resulting from the nested Fermi-surface
topology of itinerant electrons \cite{review:Mazin}. While doubts on
its validity still remain \cite{_strong:Hansmann}, this scenario
apparently does not work for the bicollinear order. On the other
hand, direct data fitting with the Heisenberg model for local spin
moments (in view of a Mott insulator \cite{_Mott:Si}) revealed
dramatic changes in the model parameters for these two orders
\cite{dft:1111:Yildirim,*dft:11:Ma,*_Mott:Fang:11}, not to mention
its difficulty to account for the metallicity.

The purpose of this Letter is to show that the unified microscopic
understanding can be achieved with a model having both components,
itinerant electrons and localized spins. It naturally possesses two
competing magnetic effects: (i) the AF superexchange coupling
$J_{ij}$ between the localized spins and (ii) the double-exchange
ferromagnetism \cite{Mn:Anderson} introduced by Hund's rule coupling
$K$ between the itinerant electrons and the localized spins.  The
competition results in the formation of antiferromagnetically
coupled ferromagnetic (FM) chains in the iron plane. These FM chains
can be \emph{straight} (Fig.~1a) or \emph{zigzag} (Fig.~1b); the
difference is small in energy but dramatic in the whole
pattern---the collinear (\textit{C}-type) or bicollinear
(\textit{E}-type) AF order. This magnetic softness is expected to
strongly scatter charge carriers above the N\'{e}el temperature,
where the system has not been frozen into a specific static order,
leading to the observed rather incoherent normal-state electron
transport \cite{syn:1111:Kamihara,trans:11:Chen,*trans:11:Hu}.

We begin with the crystal structure, which suggests that Fe$^{2+}$
is in an orbitally degenerate state, surrounded by the exceptionally
polarizable anions As, Te, or Se. As such, the Coulomb repulsion
energy $U$ between the Fe $3d$ electrons is strongly solvated,
whereas $K$ remains nearly unchanged \cite{_polar:Sawatzky}, and
this is in agreement with x-ray data on both undoped and
superconducting pnictides \cite{x:122:Yang}. Furthermore, the
solvation effect on $U$ from the anions could be strongly
orbital-dependent. A recent first-principles Wannier-function
analysis \cite{dft:1111:Lee} indicates that the influence of the
anions on the Fermi-surface states, mainly of Fe $d_{xz}$ and
$d_{yz}$ characters, is so substantial that the head-on ``$\sigma
$-bond'' hopping becomes surprisingly small and the ``$\pi $-bond''
hopping becomes the leading one.  It is likely that the $U$ for the
Fe $d_{xz}$ and $ d_{yz}$ electrons is closer to the complete
solvation. We thus assume that the Fe $3d$ electronic states
separate into two different types: The $d_{xz}$ and $d_{yz}$
electrons are itinerant and the rest form relatively localized
spins. The double-exchange FM effect is thus introduced thanks to
the energy barrier $\sim K$ for an itinerant electron to hop between
two antiparallel localized spins \cite{Mn:Anderson}; to this extent,
our proposal is supported by the spectroscopic imaging-scanning
tunneling microscopy (SI-STM) data \cite{stm:122:Chuang} on
CaFe$_{2}$As$_{2}$ and neutron-scattering data \cite{neu:11:Wen} on
Fe$_{1+\delta}$Te$_{1-x}$Se$_x$.

The minimum model considered is an effective orbital-degenerate
double-exchange model \cite{note:model}:
\begin{eqnarray}
\label{eq1} H=&-&\sum\limits_{ij\gamma \gamma^\prime \mu }
{(t_{ij}^{\gamma \gamma^\prime} C_{i\gamma \mu }^\dag C_{j
\gamma^\prime \mu}^{} +h.c.)} \nonumber \\
&-& \frac{K}{2}\sum\limits_{i\gamma \mu \mu' } {C_{i\gamma \mu
}^\dag \vec {\sigma }_{\mu \mu' } C_{i\gamma \mu' }^{} }  \cdot \vec
{S}_i + \sum\limits_{ij} { J_{ij} \vec {S}_i \cdot \vec {S}_j}
\end{eqnarray}
where $C_{i\gamma \mu }^{} $ denotes the annihilation operator of an
itinerant electron with spin $\mu=\uparrow$ or $\downarrow $ in the
$\gamma=d_{xz}$ or $d_{yz} $ orbital on site $i$. $t_{ij}^{\gamma
\gamma^\prime} $'s are the electron hopping parameters. $\vec
{\sigma }_{\mu \mu' } $ is the Pauli matrix and $\vec {S}_i$ is the
localized spin whose magnitude is $S$. $J_{ij}$ is the AF
superexchange couplings between the localized spins; in particular,
$J$ and $J'$ are respectively the nearest-neighbor (NN) and
next-nearest-neighbor (NNN) ones. $KS\simeq 0.4-0.8$ eV
\cite{x:122:Yang} and $J S^2\approx J' S^2\approx 0.01$ eV. Our
recent first-principles Wannier-function analyses on LaOFeAs
\cite{dft:1111:Lee} and FeTe suggest that to the $y$ direction, the
$d_{xz}$-$d_{xz}$ NN hopping integral $t_\| \simeq 0.4$ eV and the
$d_{yz}$-$d_{yz}$ NN hopping integral $t_\bot \simeq 0.13$ eV; they
are swapped to the $x$ direction; by symmetry the NN interorbital
hoppings are zero; the NNN intraorbital hopping integral $t'\simeq
-0.25$ eV for both $d_{xz}$ and $d_{yz}$ orbitals, and the NNN
interorbital hopping is $\pm 0.07$ eV; farther hopping parameters
and the interlayer ones are weak \cite{dft:1111:Lee} and neglected
and so are the farther superexchange parameters. We emphasize that
as demonstrated below, our conclusions are independent of the
details of the parameters as long as the following two intrinsic
features of the parameters hold: $t_\| \gg t_\bot$ and moderate $KS
\sim t_\|$. Here one itinerant electron per site (denoted as $n=1)$
is considered to correspond to the parent compounds
\cite{_oo:Kruger,note:n=3}.

For the material dependence of the parameters, note that the anion
height from the iron plane, $z_\mathrm{anion}$, is the most
significant local structural variation among the iron-based
superconductors: $z_\mathrm{anion}=1.31$, 1.35, and 1.73 {\AA} in
LaOFeAs, BaFe$_2$As$_2$, and FeTe, respectively
\cite{neu:1111:Cruz,neu:122:Huang,neu:11:Li}. Since the iron atoms
communicate with each other through the anions, the farther away the
anions are, the more isolated the iron atoms are. The isolation of
the Fe atoms would enhance the local parameters $S$ and $KS$ (in
agreement with the ordered magnetic moments of 0.36, 0.87, and 1.70
$\mu_\mathrm{B}$ in LaOFeAs, BaFe$_2$As$_2$, and FeTe, respectively
\cite{neu:1111:Cruz,neu:122:Huang,neu:11:Li}), but suppress the
nonlocal parameters $J_{ij}$. Considering the cancellation of the
$z_{\mathrm{anion}} $ effects on $S$ and $J_{ij} $, $J_{ij} S^2$ as
a whole is approximately material independent. Hence, $KS$ is
decisive in distinguishing the bicollinear ordered FeTe ($KS \sim
0.8$ eV) from the collinear ordered LaOFeAs and BaFe$_{2}$As$_{2}$
($KS \sim 0.4$ eV).

In Eq.~(1) the itinerant electrons are actually strongly correlated
via Hund's rule coupling to the quantum localized spins
\cite{Mn:Hotta,Mn:Sen}. To give a general and simple picture
elucidating that the model indeed conceives a strong magnetic phase
competition, it suffices to compare a variety of static spin orders
with the localized spins treated as Ising spins. The Ising
approximation for the $K$ term is supported by a recent numerical
study in local-density approximation plus dynamical mean-field
theory \cite{_strong:Hansmann}. Then, Eq. (\ref{eq1}) is reduced to
a system of noninteracting electrons moving in an external potential
that is $-KS/2$ and $KS/2$ at site $i$ when the itinerant electron
is spin parallel and antiparallel to $\vec{S}_i$, respectively.

The results shown in Fig.~\ref{fig:phase} indicate that a salient
feature of Eq. (\ref{eq1}) is the magnetic softness, namely the
close proximity of the collinear (\textit{C}-type), bicollinear
(\textit{E}-type), and checkerboard (\textit{G}-type) AF orders.
This key point is robust, as it exists in a quite extended
neighborhood of the realistic parameter values: $J S^2\simeq J'
S^2\simeq 0.01$ eV (Fig.~\ref{fig:phase}a and \ref{fig:phase}b),
$KS=0.4-0.8$ eV (Fig.~\ref{fig:phase}c), and $n=1$
(Fig.~\ref{fig:phase}d). It explains why a tiny change in chemical
composition can induce a dramatic change in the magnetic structure.
On the other hand, the FM (\textit{F}-type) order is shown to be a
rather high-energy and irrelevant state (Fig.~\ref{fig:phase}d).
Hence, Eq. (\ref{eq1}) with moderate $KS$ warrants the overall
in-plane AF correlations, providing a necessary environment for
forming singlet superconductivity consisting of paired electrons
with opposite spins.

Regarding the competition between the observed \textit{C}-type and
\textit{E}-type AF orders, Figs.~\ref{fig:phase}a-\ref{fig:phase}d
indicate that the \textit{C} type is favored by larger $J'S^2$,
smaller $KS$, or charge doping. Since the pure contribution of the
localized spins to the total energy per iron is
$-2S^2J'$ for the \textit{C} type and zero for the \textit{E} type,
the case where the \textit{E} type wins for $KS>0.6$ eV
(Fig.~\ref{fig:phase}c) reflects the important role of the kinetic
energy of the itinerant electrons. This is to be understood as
follows. Let us first take the heuristic limit of $KS \to \infty $
and $t_\| \gg t_\bot$: The kinetic energy per iron is $-2t_\|/\pi
\approx -0.64 t_\| $ for the \textit{C} type and $- {t_\| }\left[ {1
+ \frac{2}{\pi }\left( {\frac{{|2t'|}}{{{t_\| }}}\sin {k_{\rm{F}}} -
{k_{\rm{F}}}} \right)} \right] \approx  - 1.07{t_\| } $ (where $\cos
{k_{\rm{F}}} = |{t_\| }/2t^\prime| = 0.8$) for the \textit{E} type;
the difference is fairly enough to overcome $-2S^2J'$. Generally
speaking, the larger $KS$ is, the stronger confinement of electrons
is within the ferromagnetic chain. This, together with the strong
anisotropy in the first-neighbor hoppings, will make the electrons
in one of the $d_{xz}$ and $d_{yz}$ orbitals tend to be not
dispersive in the \textit{C} type because of its straight FM chain
structure. By contrast, in the \textit{E} type, which has zigzag FM
chains, both the $d_{xz}$ and $d_{yz}$ orbitals always equally
contribute to the kinetic energy gain. In addition, the \textit{E}
type gains the kinetic energy mainly via a Peierls-transition-type
\cite{book:Peierls} band splitting due to the doubling of the
periodicity by the alternating first-neighbor hopping strengths
($t_\| $ versus $t_\bot $; black and gray thick lines in Fig. 1b)
along the zigzag FM chain. The lower subbands are almost fully
occupied at $n=1$; thus, the \textit{E} type benefits the most near
$n=1$ and is gradually disfavored by doping, in agreement with the
neutron-scattering data \cite{neu:11:Wen,neu:11:Qiu} on
FeTe$_{1-x}$Se$_{x}$. The above analysis is robust with respect to
the electronic structure and the Fermi-surface topology because it
requires only the intrinsic (symmetry-driven) strong anisotropy in
the first-neighbor hoppings.

The \emph{metallicity} of the four typical magnetic states is
manifested in their band structures, as shown in
Fig.~\ref{fig:band}. They are presented in the momentum space
corresponding to one Fe atom per unit cell in order to explicitly
illustrate the effects of broken periodicity: Additional gap
openings and shadow bands can be clearly observed, whose intensity
reflects the strength of the bands' coupling to the orders
\cite{note:Ku}. Note that the \textit{xz} (blue) and \textit{yz}
(red) bands of the \textit{E}, \textit{G}, and \textit{F} types have
a symmetry with respect to the swap of $k_x $ and $k_y $, whereas
this symmetry is broken in the \textit{C} type, indicating an
accompanying ferro-orbital order. This agrees with our
first-principles Wannier-function analysis of LaOFeAs
\cite{dft:1111:Lee} and FeTe. A close examination of the band
structure of the \textit{C} type (Fig.~\ref{fig:band}a) reveals that
it is nearly dispersionless along the AF direction [(0,0)-($\pi
$,0)] but strongly disperses along the FM direction [(0,$\pi
)$-(0,0)], in agreement with the SI-STM measurement
\cite{stm:122:Chuang}. The same also holds for the \textit{E} type
(Fig.~\ref{fig:band}b) where the AF direction is along
($\pi$,0)-(0,$\pi$) and the FM direction along (0,0)-($\pi$,$\pi$),
and this is to be confirmed by future SI-STM measurements on FeTe.

It is interesting to point out that the zigzag view of bicollinear
is nothing but the \textit{E}-type AF order studied in the context
of $R$MnO$_3$ where $R$ is a rare-earth element \cite{Mn:Hotta}.
$R$MnO$_3$ is known to be insulating due to the
Peierls-transition-type gap opening; then, a critical question is
why the \textit{E}-type AF order of FeTe is metallic. The answer is
that the iron-based superconductors have a considerably large NNN
intraorbital hopping parameter $t^\prime$. Comparable NN and NNN
parameters are suggested by the crystal structure---the anions sit
above or below the center of the Fe plaque. Besides, that the
observed Fermi surface has a hole pocket around $(0,0)$ and an
electron pocket around $(\pi,0)$ implies that $-2t^\prime > t_\| $.
This condition is found to warrant the overlap of the split subbands
and the metallicity of the system. We verified that had
$t^\prime=0$, the \textit{E} type would be insulating. Moreover,
since the \textit{G}-type AF order (Fig. 1c) gains the kinetic
energy mainly from the $t^\prime$ term, the large $t^\prime$
introduces the \textit{G} type to the fierce phase competition.

Finally, it is noteworthy that the quantum nature of the localized
spins is important to the spin excitations, the self-energy
correction to the itinerant electrons, the electron pairing via
exchange of magnons, etc. For example, the full treatment of our
model will inevitably yield rather incoherent normal-state electron
transport.  In fact, as temperature decreases, the undoped or
underdoped compounds of the FeTe$_{1-x}$Se$_{x}$ or
FeTe$_{1-x}$S$_{x}$ family exhibit an anomalous semiconductor
behavior before getting into the metallic AF state or even the
superconducting state \cite{trans:11:Chen,trans:11:Hu}. Previously,
the large normal-state electric resistivity was used as strong
evidence for the proximity of the system to the Mott insulator like
the cuprates \cite{_Mott:Si}, while the anomaly in
FeTe$_{1-x}$Se$_{x}$ was attributed to scattering with excessive Fe
atoms \cite{trans:11:Chen,*trans:11:Hu}. The present results imply
that the phase competition among several distinct types of AF orders
is the intrinsic driving force. The relatively more severe
incoherence in FeTe$_{1-x}$Se$_{x}$ or FeTe$_{1-x}$S$_{x}$ owes its
origin to a fiercer phase competition advocated by the
\textit{E}-type AF correlation and enhanced by the Fe impurities and
the Te/Se/S disorder. Note that similar phenomena were observed in
doped manganites, where phase separation, only enhanced by quenched
disorder, has been demonstrated to be the decisive factor by using a
double-exchange model \cite{Mn:Sen} similar to Eq.~(\ref{eq1}). Our
analysis indicates that $K$ is crucial for the strongly correlated
nature of the iron-based superconductors and suggests that they are
closer kin to manganites than cuprates in terms of their magnetism
and normal-state electron transport.

In summary, we have presented an orbital-degenerate double-exchange
model that unifies the varying metallic antiferromagnetism in the
iron-based superconductors, reproducing the essential conclusions
from a number of experiments and first-principles band calculations.
The $KS$ ($z_\mathrm{anion}$) and doping induced switching of the AF
orders manifests that the sensitive competition between the
superexchange antiferromagnetism and the orbital-degenerate
double-exchange ferromagnetism is the decisive factor in the
development of magnetic correlations. Our picture is anticipated to
be instrumental for exploring other exotic properties in this new
class of superconductors such as incommensurability, the mixed
\textit{C}$_x$\textit{E}$_{1-x}$-type AF order,
electron-magnon-phonon coupling (due to the strong
$z_\mathrm{anion}$ dependence of $KS$), and ultimately the mechanism
of high-temperature superconductivity.

We thank E. Dagotto, G. Gu, J. Hill, C. C. Homes, P. D. Johnson, Q.
Li, P. B. Littlewood, C. Petrovic, M. Strongin, J. M. Tranquada, A.
M. Tsvelik, G. Xu, and I. Zaliznyak for discussions. This work was
supported by the U.S. Department of Energy (DOE), Office of Basic
Energy Science, under Contract No. DE-AC02-98CH10886, and DOE CMSN.

\textit{Note added.---}After completing the present work, we became
aware that Lv, Kr\"{u}ger, and Phillips \cite{_DE:Lv} recently used
a similar model to address the magnetic anisotropy and magnon
dispersion in the collinear AF system.


\begin{thebibliography}{34}%
\makeatletter
\providecommand \@ifxundefined [1]{%
 \@ifx{#1\undefined}
}%
\providecommand \@ifnum [1]{%
 \ifnum #1\expandafter \@firstoftwo
 \else \expandafter \@secondoftwo
 \fi
}%
\providecommand \@ifx [1]{%
 \ifx #1\expandafter \@firstoftwo
 \else \expandafter \@secondoftwo
 \fi
}%
\providecommand \natexlab [1]{#1}%
\providecommand \enquote  [1]{``#1''}%
\providecommand \bibnamefont  [1]{#1}%
\providecommand \bibfnamefont [1]{#1}%
\providecommand \citenamefont [1]{#1}%
\providecommand \href@noop [0]{\@secondoftwo}%
\providecommand \href [0]{\begingroup \@sanitize@url \@href}%
\providecommand \@href[1]{\@@startlink{#1}\@@href}%
\providecommand \@@href[1]{\endgroup#1\@@endlink}%
\providecommand \@sanitize@url [0]{\catcode `\\12\catcode
`\$12\catcode
  `\&12\catcode `\#12\catcode `\^12\catcode `\_12\catcode `\%12\relax}%
\providecommand \@@startlink[1]{}%
\providecommand \@@endlink[0]{}%
\providecommand \url  [0]{\begingroup\@sanitize@url \@url }%
\providecommand \@url [1]{\endgroup\@href {#1}{\urlprefix }}%
\providecommand \urlprefix  [0]{URL }%
\providecommand \Eprint [0]{\href }%
\@ifxundefined \urlstyle {%
  \providecommand \doi  [0]{\begingroup \@sanitize@url \@doi}%
  \providecommand \@doi [1]{\endgroup \@@startlink {\doibase
  #1}doi:\discretionary {}{}{}#1\@@endlink }%
}{%
  \providecommand \doi  [0]{doi:\discretionary{}{}{}\begingroup
  \urlstyle{rm}\Url }%
}%
\providecommand \doibase [0]{http://dx.doi.org/}%
\providecommand \Doi [0]{\begingroup \@sanitize@url \@Doi }%
\providecommand \@Doi  [1]{\endgroup\@@startlink{\doibase#1}\@@Doi}%
\providecommand \@@Doi [1]{#1\@@endlink}%
\providecommand \selectlanguage [0]{\@gobble}%
\providecommand \bibinfo  [0]{\@secondoftwo}%
\providecommand \bibfield  [0]{\@secondoftwo}%
\providecommand \translation [1]{[#1]}%
\providecommand \BibitemOpen [0]{}%
\providecommand \bibitemStop [0]{}%
\providecommand \bibitemNoStop [0]{.\EOS\space}%
\providecommand \EOS [0]{\spacefactor3000\relax}%
\providecommand \BibitemShut  [1]{\csname bibitem#1\endcsname}%
\bibitem [{\citenamefont {Kamihara}\ \emph {et~al.}(2008)\citenamefont
  {Kamihara}, \citenamefont {Watanabe}, \citenamefont {Hirano},\ and\
  \citenamefont {Hosono}}]{syn:1111:Kamihara}%
  \BibitemOpen
  \bibfield  {author} {\bibinfo {author} {\bibfnamefont {Y.}~\bibnamefont
  {Kamihara}}, \bibinfo {author} {\bibfnamefont {T.}~\bibnamefont {Watanabe}},
  \bibinfo {author} {\bibfnamefont {M.}~\bibnamefont {Hirano}}, \ and\ \bibinfo
  {author} {\bibfnamefont {H.}~\bibnamefont {Hosono}},\ }\href@noop {}
  {\bibfield  {journal} {\bibinfo  {journal} {J. Am. Chem. Soc.},\ }\textbf
  {\bibinfo {volume} {130}},\ \bibinfo {pages} {3296} (\bibinfo {year}
  {2008})}\BibitemShut {NoStop}%
\bibitem [{\citenamefont {Rotter}\ \emph {et~al.}(2008)\citenamefont {Rotter},
  \citenamefont {Tegel},\ and\ \citenamefont {Johrendt}}]{syn:122:Rotter}%
  \BibitemOpen
  \bibfield  {author} {\bibinfo {author} {\bibfnamefont {M.}~\bibnamefont
  {Rotter}}, \bibinfo {author} {\bibfnamefont {M.}~\bibnamefont {Tegel}}, \
  and\ \bibinfo {author} {\bibfnamefont {D.}~\bibnamefont {Johrendt}},\
  }\href@noop {} {\bibfield  {journal} {\bibinfo  {journal} {Phys. Rev.
  Lett.},\ }\textbf {\bibinfo {volume} {101}},\ \bibinfo {pages} {107006}
  (\bibinfo {year} {2008})}\BibitemShut {NoStop}%
\bibitem [{\citenamefont {\textit{et al.}}(2008){\natexlab{a}}}]{syn:11:Yeh}%
  \BibitemOpen
  \bibfield  {author} {\bibinfo {author} {\bibfnamefont {K.-W.~Yeh}\
  \bibnamefont {\textit{et al.}}},\ }\href@noop {} {\bibfield  {journal}
  {\bibinfo  {journal} {Europhys. Lett.},\ }\textbf {\bibinfo {volume} {84}},\
  \bibinfo {pages} {37002} (\bibinfo {year} {2008}{\natexlab{a}})}\BibitemShut
  {NoStop}%
\bibitem [{\citenamefont {de~la Cruz}\ \emph {et~al.}(2008)\citenamefont {de~la
  Cruz}, \citenamefont {Huang}, \citenamefont {Lynn}, \citenamefont {Li},
  \citenamefont {II}, \citenamefont {Zarestky}, \citenamefont {Mook},
  \citenamefont {Chen}, \citenamefont {Luo}, \citenamefont {Wang},\ and\
  \citenamefont {Dai}}]{neu:1111:Cruz}%
  \BibitemOpen
  \bibfield  {author} {\bibinfo {author} {\bibfnamefont {C.}~\bibnamefont
  {de~la Cruz \etal}},\ }\href@noop {}
  {\bibfield  {journal} {\bibinfo  {journal} {Nature},\ }\textbf {\bibinfo
  {volume} {453}},\ \bibinfo {pages} {899} (\bibinfo {year}
  {2008})}\BibitemShut {NoStop}%
\bibitem [{\citenamefont {Huang}\ \emph {et~al.}(2008)\citenamefont {Huang},
  \citenamefont {Qiu}, \citenamefont {Bao}, \citenamefont {Green},
  \citenamefont {Lynn}, \citenamefont {Gasparovic}, \citenamefont {Wu},
  \citenamefont {Wu},\ and\ \citenamefont {Chen}}]{neu:122:Huang}%
  \BibitemOpen
  \bibfield  {author} {\bibinfo {author} {\bibfnamefont {Q.}~\bibnamefont
  {Huang \etal}},\ }\href@noop {} {\bibfield  {journal} {\bibinfo
  {journal} {Phys. Rev. Lett.},\ }\textbf {\bibinfo {volume} {101}},\ \bibinfo
  {pages} {257003} (\bibinfo {year} {2008})}\BibitemShut {NoStop}%
\bibitem [{\citenamefont {Li}\ \emph {et~al.}(2009)\citenamefont {Li},
  \citenamefont {de~la Cruz}, \citenamefont {Huang}, \citenamefont {Chen},
  \citenamefont {Lynn}, \citenamefont {Hu}, \citenamefont {Huang},
  \citenamefont {Hsu}, \citenamefont {Yeh}, \citenamefont {Wu},\ and\
  \citenamefont {Dai}}]{neu:11:Li}%
  \BibitemOpen
  \bibfield  {author} {\bibinfo {author} {\bibfnamefont {S.}~\bibnamefont
  {Li \etal}},\ }\href@noop {}
  {\bibfield  {journal} {\bibinfo  {journal} {Phys. Rev. B},\ }\textbf
  {\bibinfo {volume} {79}},\ \bibinfo {pages} {054503} (\bibinfo {year}
  {2009})}\BibitemShut {NoStop}%
\bibitem [{\citenamefont {Christianson}\ \emph {et~al.}(2008)\citenamefont
  {Christianson}, \citenamefont {Goremychkin}, \citenamefont {Osborn},
  \citenamefont {Rosenkranz}, \citenamefont {Lumsden}, \citenamefont
  {Malliakas}, \citenamefont {Todorov}, \citenamefont {Claus}, \citenamefont
  {Chung}, \citenamefont {Kanatzidis}, \citenamefont {Bewley},\ and\
  \citenamefont {Guidi}}]{neu:122:Christianson}%
  \BibitemOpen
  \bibfield  {author} {\bibinfo {author} {\bibfnamefont {A.~D.}\ \bibnamefont
  {Christianson \etal}},\ }\href@noop {} {\bibfield  {journal} {\bibinfo
  {journal} {Nature},\ }\textbf {\bibinfo {volume} {456}},\ \bibinfo {pages}
  {930} (\bibinfo {year} {2008})}\BibitemShut {NoStop}%
\bibitem [{\citenamefont {Qiu}\ \emph {et~al.}(2009)\citenamefont {Qiu},
  \citenamefont {Bao}, \citenamefont {Zhao}, \citenamefont {Broholm},
  \citenamefont {Stanev}, \citenamefont {Tesanovic}, \citenamefont
  {Gasparovic}, \citenamefont {Chang}, \citenamefont {Hu}, \citenamefont
  {Qian}, \citenamefont {Fang},\ and\ \citenamefont {Mao}}]{neu:11:Qiu}%
  \BibitemOpen
  \bibfield  {author} {\bibinfo {author} {\bibfnamefont {Y.}~\bibnamefont
  {Qiu \etal}},\ }\href@noop {}
  {\bibfield  {journal} {\bibinfo  {journal} {Phys. Rev. Lett.},\ }\textbf
  {\bibinfo {volume} {103}},\ \bibinfo {pages} {067008} (\bibinfo {year}
  {2009})}\BibitemShut {NoStop}%
\bibitem [{\citenamefont {Mazin}(2010)}]{review:Mazin}%
  \BibitemOpen
  \bibfield  {author} {\bibinfo {author} {\bibfnamefont {I.~I.}\ \bibnamefont
  {Mazin}},\ }\href@noop {} {\bibfield  {journal} {\bibinfo  {journal}
  {Nature},\ }\textbf {\bibinfo {volume} {464}},\ \bibinfo {pages} {183}
  (\bibinfo {year} {2010})}\BibitemShut {NoStop}%
\bibitem [{\citenamefont {Lumsden}\ and\ \citenamefont
  {Christianson}(2010)}]{review:Lumsden}%
  \BibitemOpen
  \bibfield  {author} {\bibinfo {author} {\bibfnamefont {M.}~\bibnamefont
  {Lumsden}}\ and\ \bibinfo {author} {\bibfnamefont {A.}~\bibnamefont
  {Christianson}},\ }\href@noop {} {\bibfield  {journal} {\bibinfo  {journal}
  {J. Phys.: Condens. Matter},\ }\textbf {\bibinfo {volume} {22}},\ \bibinfo
  {pages} {203203} (\bibinfo {year} {2010})}\BibitemShut {NoStop}%
\bibitem [{\citenamefont {\textit{et
  al.}}(2008){\natexlab{b}}}]{arpes:1111:Lu}%
  \BibitemOpen
  \bibfield  {author} {\bibinfo {author} {\bibfnamefont {D.~H.~Lu}\
  \bibnamefont {\textit{et al.}}},\ }\href@noop {} {\bibfield  {journal}
  {\bibinfo  {journal} {Nature},\ }\textbf {\bibinfo {volume} {455}},\ \bibinfo
  {pages} {81} (\bibinfo {year} {2008}{\natexlab{b}})}\BibitemShut {NoStop}%
\bibitem [{\citenamefont {\textit{et
  al.}}(2008){\natexlab{c}}}]{arpes:122:Ding}%
  \BibitemOpen
  \bibfield  {author} {\bibinfo {author} {\bibfnamefont {H.~Ding}\ \bibnamefont
  {\textit{et al.}}},\ }\href@noop {} {\bibfield  {journal} {\bibinfo
  {journal} {Europhys. Lett.},\ }\textbf {\bibinfo {volume} {83}},\ \bibinfo
  {pages} {47001} (\bibinfo {year} {2008}{\natexlab{c}})}\BibitemShut {NoStop}%
\bibitem [{\citenamefont {Xia}\ \emph {et~al.}(2009)\citenamefont {Xia},
  \citenamefont {Qian}, \citenamefont {Wray}, \citenamefont {Hsieh},
  \citenamefont {Chen}, \citenamefont {Luo}, \citenamefont {Wang},\ and\
  \citenamefont {Hasan}}]{arpes:11:Xia}%
  \BibitemOpen
  \bibfield  {author} {\bibinfo {author} {\bibfnamefont {Y.}~\bibnamefont
  {Xia \etal}},\ }\href@noop {} {\bibfield  {journal} {\bibinfo  {journal} {Phys.
  Rev. Lett.},\ }\textbf {\bibinfo {volume} {103}},\ \bibinfo {pages} {037002}
  (\bibinfo {year} {2009})}\BibitemShut {NoStop}%
\bibitem [{\citenamefont {Moon}\ and\ \citenamefont
  {Choi}(2010)}]{dft:11:Moon}%
  \BibitemOpen
  \bibfield  {author} {\bibinfo {author} {\bibfnamefont {C.-Y.}\ \bibnamefont
  {Moon}}\ and\ \bibinfo {author} {\bibfnamefont {H.~J.}\ \bibnamefont
  {Choi}},\ }\href@noop {} {\bibfield  {journal} {\bibinfo  {journal} {Phys.
  Rev. Lett.},\ }\textbf {\bibinfo {volume} {104}},\ \bibinfo {pages} {057003}
  (\bibinfo {year} {2010})}\BibitemShut {NoStop}%
\bibitem [{\citenamefont {Hansmann}\ \emph {et~al.}(2010)\citenamefont
  {Hansmann}, \citenamefont {Arita}, \citenamefont {Toschi}, \citenamefont
  {Sakai}, \citenamefont {Sangiovanni},\ and\ \citenamefont
  {Held}}]{_strong:Hansmann}%
  \BibitemOpen
  \bibfield  {author} {\bibinfo {author} {\bibfnamefont {P.}~\bibnamefont
  {Hansmann}}, \bibinfo {author} {\bibfnamefont {R.}~\bibnamefont {Arita}},
  \bibinfo {author} {\bibfnamefont {A.}~\bibnamefont {Toschi}}, \bibinfo
  {author} {\bibfnamefont {S.}~\bibnamefont {Sakai}}, \bibinfo {author}
  {\bibfnamefont {G.}~\bibnamefont {Sangiovanni}}, \ and\ \bibinfo {author}
  {\bibfnamefont {K.}~\bibnamefont {Held}},\ }\href@noop {} {\bibfield
  {journal} {\bibinfo  {journal} {Phys. Rev. Lett.},\ }\textbf {\bibinfo
  {volume} {104}},\ \bibinfo {pages} {197002} (\bibinfo {year}
  {2010})}\BibitemShut {NoStop}%
\bibitem [{\citenamefont {Si}\ and\ \citenamefont {Abrahams}(2008)}]{_Mott:Si}%
  \BibitemOpen
  \bibfield  {author} {\bibinfo {author} {\bibfnamefont {Q.}~\bibnamefont
  {Si}}\ and\ \bibinfo {author} {\bibfnamefont {E.}~\bibnamefont {Abrahams}},\
  }\href@noop {} {\bibfield  {journal} {\bibinfo  {journal} {Phys. Rev.
  Lett.},\ }\textbf {\bibinfo {volume} {101}},\ \bibinfo {pages} {076401}
  (\bibinfo {year} {2008})}\BibitemShut {NoStop}%
\bibitem [{\citenamefont {Yildirim}(2008)}]{dft:1111:Yildirim}%
  \BibitemOpen
  \bibfield  {author} {\bibinfo {author} {\bibfnamefont {T.}~\bibnamefont
  {Yildirim}},\ }\href@noop {} {\bibfield  {journal} {\bibinfo  {journal}
  {Phys. Rev. Lett.},\ }\textbf {\bibinfo {volume} {101}},\ \bibinfo {pages}
  {057010} (\bibinfo {year} {2008})}\BibitemShut {NoStop}%
\bibitem [{\citenamefont {Ma}\ \emph {et~al.}(2009)\citenamefont {Ma},
  \citenamefont {Ji}, \citenamefont {Hu}, \citenamefont {Lu},\ and\
  \citenamefont {Xiang}}]{dft:11:Ma}%
  \BibitemOpen
  \bibfield  {author} {\bibinfo {author} {\bibfnamefont {F.}~\bibnamefont
  {Ma}}, \bibinfo {author} {\bibfnamefont {W.}~\bibnamefont {Ji}}, \bibinfo
  {author} {\bibfnamefont {J.}~\bibnamefont {Hu}}, \bibinfo {author}
  {\bibfnamefont {Z.-Y.}\ \bibnamefont {Lu}}, \ and\ \bibinfo {author}
  {\bibfnamefont {T.}~\bibnamefont {Xiang}},\ }\href@noop {} {\bibfield
  {journal} {\bibinfo  {journal} {Phys. Rev. Lett.},\ }\textbf {\bibinfo
  {volume} {102}},\ \bibinfo {pages} {177003} (\bibinfo {year}
  {2009})}\BibitemShut {NoStop}%
\bibitem [{\citenamefont {Fang}\ \emph {et~al.}(2009)\citenamefont {Fang},
  \citenamefont {Bernevig},\ and\ \citenamefont {Hu}}]{_Mott:Fang:11}%
  \BibitemOpen
  \bibfield  {author} {\bibinfo {author} {\bibfnamefont {C.}~\bibnamefont
  {Fang}}, \bibinfo {author} {\bibfnamefont {B.~A.}\ \bibnamefont {Bernevig}},
  \ and\ \bibinfo {author} {\bibfnamefont {J.}~\bibnamefont {Hu}},\ }\href@noop
  {} {\bibfield  {journal} {\bibinfo  {journal} {Europhys. Lett.},\ }\textbf
  {\bibinfo {volume} {86}},\ \bibinfo {pages} {67005} (\bibinfo {year}
  {2009})}\BibitemShut {NoStop}%
\bibitem [{\citenamefont {Anderson}(1955)}]{Mn:Anderson}%
  \BibitemOpen
  \bibfield  {author} {\bibinfo {author} {\bibfnamefont {P.~W.}\ \bibnamefont
  {Anderson}},\ }\href@noop {} {\bibfield  {journal} {\bibinfo  {journal}
  {Phys. Rev.},\ }\textbf {\bibinfo {volume} {100}},\ \bibinfo {pages} {675}
  (\bibinfo {year} {1955})}\BibitemShut {NoStop}%
\bibitem [{\citenamefont {Chen}\ \emph {et~al.}(2009)\citenamefont {Chen},
  \citenamefont {Chen}, \citenamefont {Dong}, \citenamefont {Hu}, \citenamefont
  {Li}, \citenamefont {Zhang}, \citenamefont {Zheng}, \citenamefont {Luo},\
  and\ \citenamefont {Wang}}]{trans:11:Chen}%
  \BibitemOpen
  \bibfield  {author} {\bibinfo {author} {\bibfnamefont {G.~F.}\ \bibnamefont
  {Chen \etal}},\
  }\href@noop {} {\bibfield  {journal} {\bibinfo  {journal} {Phys. Rev. B},\
  }\textbf {\bibinfo {volume} {79}},\ \bibinfo {pages} {140509(R)} (\bibinfo
  {year} {2009})}\BibitemShut {NoStop}%
\bibitem [{\citenamefont {Hu}\ \emph {et~al.}(2009)\citenamefont {Hu},
  \citenamefont {Bozin}, \citenamefont {Warren},\ and\ \citenamefont
  {Petrovic}}]{trans:11:Hu}%
  \BibitemOpen
  \bibfield  {author} {\bibinfo {author} {\bibfnamefont {R.}~\bibnamefont
  {Hu}}, \bibinfo {author} {\bibfnamefont {E.~S.}\ \bibnamefont {Bozin}},
  \bibinfo {author} {\bibfnamefont {J.~B.}\ \bibnamefont {Warren}}, \ and\
  \bibinfo {author} {\bibfnamefont {C.}~\bibnamefont {Petrovic}},\ }\href@noop
  {} {\bibfield  {journal} {\bibinfo  {journal} {Phys. Rev. B},\ }\textbf
  {\bibinfo {volume} {80}},\ \bibinfo {pages} {214514} (\bibinfo {year}
  {2009})}\BibitemShut {NoStop}%
\bibitem [{\citenamefont {Sawatzky}\ \emph {et~al.}(2009)\citenamefont
  {Sawatzky}, \citenamefont {Elfimov}, \citenamefont {van~den Brink},\ and\
  \citenamefont {Zaanen}}]{_polar:Sawatzky}%
  \BibitemOpen
  \bibfield  {author} {\bibinfo {author} {\bibfnamefont {G.~A.}\ \bibnamefont
  {Sawatzky}}, \bibinfo {author} {\bibfnamefont {I.~S.}\ \bibnamefont
  {Elfimov}}, \bibinfo {author} {\bibfnamefont {J.}~\bibnamefont {van~den
  Brink}}, \ and\ \bibinfo {author} {\bibfnamefont {J.}~\bibnamefont
  {Zaanen}},\ }\href@noop {} {\bibfield  {journal} {\bibinfo  {journal}
  {Europhys. Lett.},\ }\textbf {\bibinfo {volume} {86}},\ \bibinfo {pages}
  {17006} (\bibinfo {year} {2009})}\BibitemShut {NoStop}%
\bibitem [{\citenamefont {Yang}\ \emph {et~al.}(2009)\citenamefont {Yang},
  \citenamefont {Sorini}, \citenamefont {Chen}, \citenamefont {Moritz},
  \citenamefont {Lee}, \citenamefont {Vernay}, \citenamefont {Olalde-Velasco},
  \citenamefont {Denlinger}, \citenamefont {Delley}, \citenamefont {Chu},
  \citenamefont {Analytis}, \citenamefont {Fisher}, \citenamefont {Ren},
  \citenamefont {Yang}, \citenamefont {Lu}, \citenamefont {Zhao}, \citenamefont
  {van~den Brink}, \citenamefont {Hussain}, \citenamefont {Shen},\ and\
  \citenamefont {Devereaux}}]{x:122:Yang}%
  \BibitemOpen
  \bibfield  {author} {\bibinfo {author} {\bibfnamefont {W.~L.}\ \bibnamefont
  {Yang \etal}},\ }\href@noop {} {\bibfield  {journal} {\bibinfo
  {journal} {Phys. Rev. B},\ }\textbf {\bibinfo {volume} {80}},\ \bibinfo
  {pages} {014508} (\bibinfo {year} {2009})}\BibitemShut {NoStop}%
\bibitem [{\citenamefont {Lee}\ \emph {et~al.}(2009)\citenamefont {Lee},
  \citenamefont {Yin},\ and\ \citenamefont {Ku}}]{dft:1111:Lee}%
  \BibitemOpen
  \bibfield  {author} {\bibinfo {author} {\bibfnamefont {C.-C.}\ \bibnamefont
  {Lee}}, \bibinfo {author} {\bibfnamefont {W.-G.}\ \bibnamefont {Yin}}, \ and\
  \bibinfo {author} {\bibfnamefont {W.}~\bibnamefont {Ku}},\ }\href@noop {}
  {\bibfield  {journal} {\bibinfo  {journal} {Phys. Rev. Lett.},\ }\textbf
  {\bibinfo {volume} {103}},\ \bibinfo {pages} {267001} (\bibinfo {year}
  {2009})}\BibitemShut {NoStop}%
\bibitem [{\citenamefont {Chuang}\ \emph {et~al.}(2010)\citenamefont {Chuang},
  \citenamefont {Allan}, \citenamefont {Lee}, \citenamefont {Xie},
  \citenamefont {Ni}, \citenamefont {Bud'ko}, \citenamefont {Boebinger},
  \citenamefont {Canfield},\ and\ \citenamefont {Davis}}]{stm:122:Chuang}%
  \BibitemOpen
  \bibfield  {author} {\bibinfo {author} {\bibfnamefont {T.-M.}\ \bibnamefont
  {Chuang \etal}},\ }\href@noop {} {\bibfield  {journal} {\bibinfo
  {journal} {Science},\ }\textbf {\bibinfo {volume} {327}},\ \bibinfo {pages}
  {181} (\bibinfo {year} {2010})}\BibitemShut {NoStop}%
\bibitem [{\citenamefont {Wen}\ \emph {et~al.}(2009)\citenamefont {Wen},
  \citenamefont {Xu}, \citenamefont {Xu}, \citenamefont {Lin}, \citenamefont
  {Li}, \citenamefont {Ratcliff}, \citenamefont {Gu},\ and\ \citenamefont
  {Tranquada}}]{neu:11:Wen}%
  \BibitemOpen
  \bibfield  {author} {\bibinfo {author} {\bibfnamefont {J.}~\bibnamefont
  {Wen \etal}},\
  }\href@noop {} {\bibfield  {journal} {\bibinfo  {journal} {Phys. Rev. B},\
  }\textbf {\bibinfo {volume} {80}},\ \bibinfo {pages} {104506} (\bibinfo
  {year} {2009})}\BibitemShut {NoStop}%
\bibitem [{not(){\natexlab{a}}}]{note:model}%
  \BibitemOpen
  \href@noop {} {} {\natexlab{a}} \bibinfo {note} {One may include the
  solvated interaction between the itinerant electrons for a more sophisticate
  model.}\BibitemShut {Stop}%
\bibitem [{\citenamefont {Kr{\"{u}}ger}\ \emph {et~al.}(2009)\citenamefont
  {Kr{\"{u}}ger}, \citenamefont {Kumar}, \citenamefont {Zaanen},\ and\
  \citenamefont {van~den Brink}}]{_oo:Kruger}%
  \BibitemOpen
  \bibfield  {author} {\bibinfo {author} {\bibfnamefont {F.}~\bibnamefont
  {Kr{\"{u}}ger}}, \bibinfo {author} {\bibfnamefont {S.}~\bibnamefont {Kumar}},
  \bibinfo {author} {\bibfnamefont {J.}~\bibnamefont {Zaanen}}, \ and\ \bibinfo
  {author} {\bibfnamefont {J.}~\bibnamefont {van~den Brink}},\ }\href@noop {}
  {\bibfield  {journal} {\bibinfo  {journal} {Phys. Rev. B},\ }\textbf
  {\bibinfo {volume} {79}},\ \bibinfo {pages} {054504} (\bibinfo {year}
  {2009})}\BibitemShut {NoStop}%
\bibitem [{not(){\natexlab{b}}}]{note:n=3}%
  \BibitemOpen
  \href@noop {} {} {\natexlab{b}} \bibinfo {note} {This corresponds to the
  intermediate-spin configuration. With $t_\| \gg t_\bot$ and moderate $KS$ our
  conclusions also apply to the case of the one hole per site ($n=3$) for the
  high-spin configuration \cite{dft:1111:Lee}.}\BibitemShut {Stop}%
\bibitem [{\citenamefont {Hotta}\ \emph {et~al.}(2003)\citenamefont {Hotta},
  \citenamefont {Moraghebi}, \citenamefont {Feiguin}, \citenamefont {Moreo},
  \citenamefont {Yunoki},\ and\ \citenamefont {Dagotto}}]{Mn:Hotta}%
  \BibitemOpen
  \bibfield  {author} {\bibinfo {author} {\bibfnamefont {T.}~\bibnamefont
  {Hotta}}, \bibinfo {author} {\bibfnamefont {M.}~\bibnamefont {Moraghebi}},
  \bibinfo {author} {\bibfnamefont {A.}~\bibnamefont {Feiguin}}, \bibinfo
  {author} {\bibfnamefont {A.}~\bibnamefont {Moreo}}, \bibinfo {author}
  {\bibfnamefont {S.}~\bibnamefont {Yunoki}}, \ and\ \bibinfo {author}
  {\bibfnamefont {E.}~\bibnamefont {Dagotto}},\ }\href@noop {} {\bibfield
  {journal} {\bibinfo  {journal} {Phys. Rev. Lett.},\ }\textbf {\bibinfo
  {volume} {90}},\ \bibinfo {pages} {247203} (\bibinfo {year}
  {2003})}\BibitemShut {NoStop}%
\bibitem [{\citenamefont {\c{S}en}\ \emph {et~al.}(2007)\citenamefont
  {\c{S}en}, \citenamefont {Alvarez},\ and\ \citenamefont {Dagotto}}]{Mn:Sen}%
  \BibitemOpen
  \bibfield  {author} {\bibinfo {author} {\bibfnamefont {C.}~\bibnamefont
  {\c{S}en}}, \bibinfo {author} {\bibfnamefont {G.}~\bibnamefont {Alvarez}}, \
  and\ \bibinfo {author} {\bibfnamefont {E.}~\bibnamefont {Dagotto}},\
  }\href@noop {} {\bibfield  {journal} {\bibinfo  {journal} {Phys. Rev.
  Lett.},\ }\textbf {\bibinfo {volume} {98}},\ \bibinfo {pages} {127202}
  (\bibinfo {year} {2007})}\BibitemShut {NoStop}%
\bibitem [{\citenamefont {Peierls}(1955)}]{book:Peierls}%
  \BibitemOpen
  \bibfield  {author} {\bibinfo {author} {\bibfnamefont {R.~E.}\ \bibnamefont
  {Peierls}},\ }\href@noop {} {\emph {\bibinfo {title} {Quantum Theory of
  Solids}}}\ (\bibinfo  {publisher} {Oxford Univiversity Press},\ \bibinfo {address}
  {Oxford, England},\ \bibinfo {year} {1955})\ p.\ \bibinfo {pages} {108},\
  \bibinfo {note} {p. 108}\BibitemShut {NoStop}%
\bibitem [{\citenamefont {Ku}\ \emph {et~al.}(2010)\citenamefont {Ku},
  \citenamefont {Berlijn},\ and\ \citenamefont {Lee}}]{note:Ku}%
  \BibitemOpen
  \bibfield  {author} {\bibinfo {author} {\bibfnamefont {W.}~\bibnamefont
  {Ku}}, \bibinfo {author} {\bibfnamefont {T.}~\bibnamefont {Berlijn}}, \ and\
  \bibinfo {author} {\bibfnamefont {C.-C.}\ \bibnamefont {Lee}},\ }\href@noop
  {} {\bibfield  {journal} {\bibinfo  {journal} {Phys. Rev. Lett.},\ }\textbf
  {\bibinfo {volume} {104}},\ \bibinfo {pages} {216401} (\bibinfo {year}
  {2010})}\BibitemShut {NoStop}%
\bibitem{_DE:Lv} W. Lv, F. Kr\"{u}ger, and P. Phillips,
Phys. Rev. B \textbf{82}, 045125 (2010).
\end{thebibliography}

%

\begin{figure*}[b]
\vspace{1cm}
\includegraphics[width=0.35\columnwidth,clip=true,angle=270]{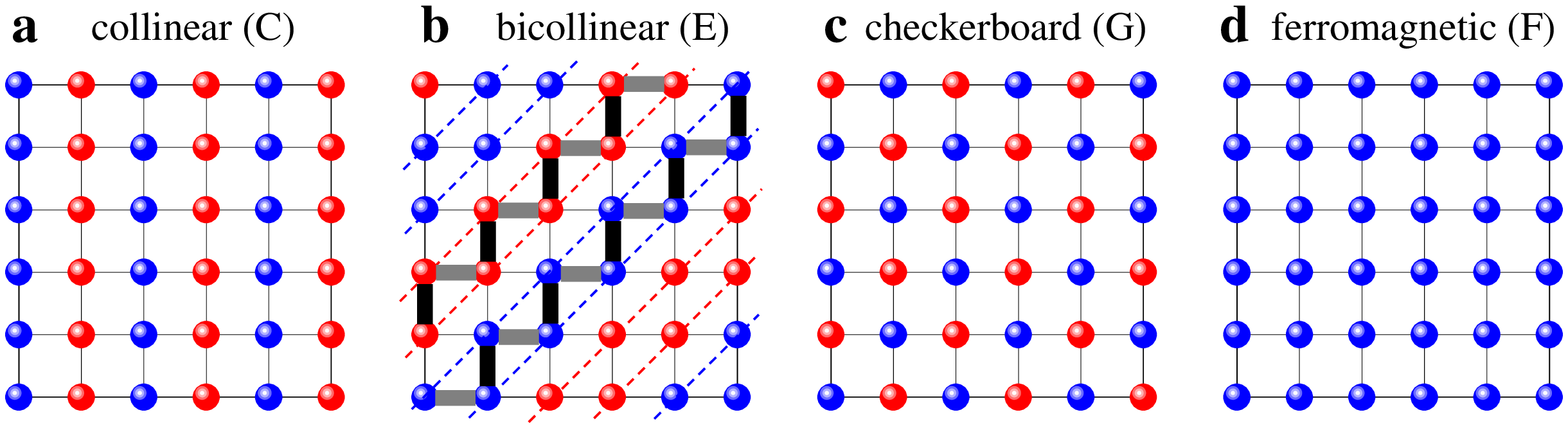}
\caption{\label{fig1} The in-plane patterns of the spin-up (blue
balls) and spin-down (red balls) iron atoms in (a) the collinear
(\textit{C}-type), (b) bicollinear (\textit{E}-type), and (c)
checkerboard (\textit{G}-type) AF orders. Note that bicollinear
means to follow the dashed lines for FM correlation, while a more
insightful view is to follow the \emph{zigzag} thick lines (black
and gray stand for alternating electron hopping strengths) of the
\textit{E} type.
}
\end{figure*}

\begin{figure*}[t]
\vspace{1cm}
\includegraphics[width=0.65\columnwidth,clip=true,angle=270]{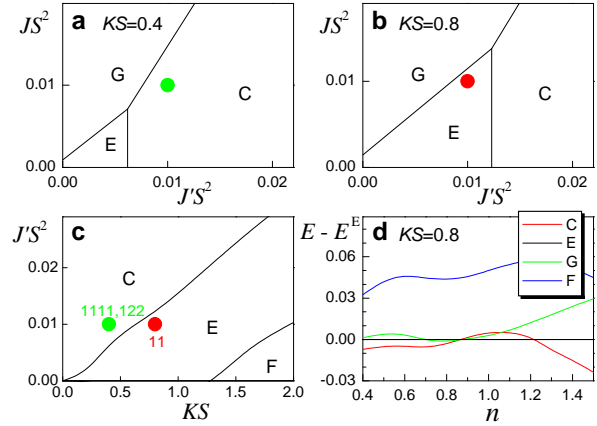}
\caption{\label{fig:phase} \textbf{Close proximity of AF orders.}
The shorthand notations \cite{Mn:Hotta} are \textit{G}
(checkerboard), \textit{C} (collinear), \textit{E} (bicollinear),
and \textit{F} (ferromagnetic). The $J S^2-J' S^2$ phase diagrams
for $n=1$ with (a) $KS=0.4$ eV and (b) $KS=0.8$ eV. The green and
red dots mark out $J S^2=J' S^2=0.01$ eV. (c) The $J' S^2-KS$ phase
diagrams for $n=1$. Also illustrated are the placements of FeTe
(``11'', red dot), LaOFeAs and BaFe$_{2}$As$_{2}$ (``1111'' and
``122'', respectively, green dot). $J S^2=J' S^2$. (d) The total
energy as a function of $n$ with respect to that of the bicollinear
order. $KS=0.8$ eV and $J S^2=J' S^2=0.01$ eV.
The energy unit is eV.}
\end{figure*}

\begin{figure*}[t]
\vspace{1cm}
\includegraphics[width=0.95\columnwidth,clip=true,angle=270]{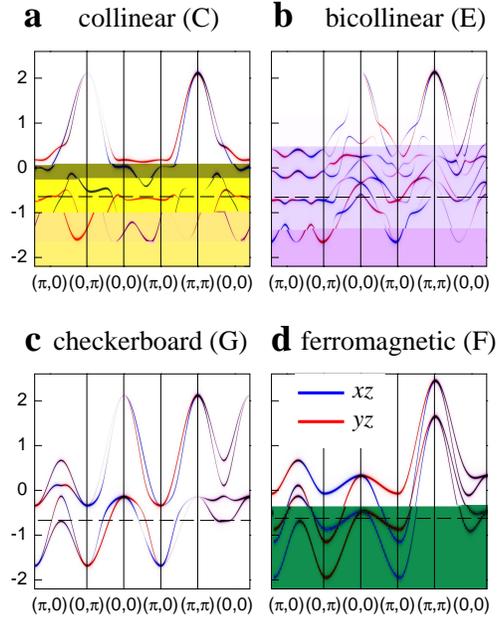}
\caption{\label{fig:band} Electronic structures of the itinerant
electrons for the (a) collinear, (b) bicollinear, (c) checkerboard,
and (d) ferromagnetic spin orders calculated with $KS=0.8$ eV. The
dashed lines are the Fermi level for $n=1$. The energy unit is eV.}
\end{figure*}

\end{document}